# MULTIPLE CHARGE STATE BEAM ACCELERATION AT ATLAS

P.N. Ostroumov, R.C. Pardo, G.P. Zinkann, K.W. Shepard, J.A. Nolen, Physics Division, ANL,
9700 S. Cass Avenue, Argonne, IL60439, USA

*Abstract*

A test of the acceleration of multiple charge-state uranium beams was performed at the ATLAS accelerator. A $^{238}U^{+26}$ beam was accelerated in the ATLAS PII linac to 286 MeV (~1.2 MeV/u) and stripped in a carbon foil located 0.5 m from the entrance of the ATLAS Booster section. A $^{58}Ni^{9+}$ 'guide' beam from the tandem injector was used to tune the Booster for $^{238}U^{+38}$. All charge states from the stripping were injected into the booster and accelerated. Up to 94% of the beam was accelerated through the Booster linac, with losses mostly in the lower charge states. The measured beam properties of each charge state and a comparison to numerical simulations are reported in this paper.

## 1 INTRODUCTION

Simultaneous acceleration of multiple charge-state beams has been proposed as a method of substantially increasing the available beam current for the heaviest ions from a RIA (Rare Isotope Accelerator) driver linac [1]. There is presently no facility where multiple charge-state beam acceleration is used to increase the beam current. Therefore, in order to demonstrate the concept, we have accelerated a multiple charge-state uranium beam in the existing ATLAS heavy-ion linac, and performed careful measurements of the accelerated beam parameters for comparison with the results of numerical simulations.

The acceleration of multiple charge-state uranium beams has been observed at the ATLAS 'booster' as part of the 'normal' uranium beam configuration. However, the multiple charge states have been considered parasitic. Therefore systematic studies of all the accelerated charge states were not performed and accelerator parameters were not chosen to optimize the acceleration of the other charge states.

In this test, a $^{238}U^{+26}$ beam was accelerated to 286 MeV (~1.2 MeV/u) and stripped. All charge states near $q_0$=38 were then simultaneously accelerated in the ATLAS 'Booster' linac. The parameters of each selected charge state were carefully measured.

## 2 BEAM DYNAMICS SIMULATIONS

For a better understanding of the beam test results, a multiple charge-state beam dynamics simulation in the Booster was performed with the modified LANA code [2]. At the position of the stripping target, input beam parameters were assumed to be the same for all charge states. The beam longitudinal and transverse emittances were taken to be equal to $\varepsilon_L$=2 $\pi$·keV/u·nsec and $\varepsilon_T$=0.25 $\pi$·mm·mrad. The ray-tracing code incorporated actual resonator field profiles and field levels of the superconducting cavities. The synchronous phase for $^{238}U^{+38}$ was set to -30°. Fig. 1 and 2 show the calculated longitudinal phase space at the booster exit and the transverse beam envelopes along the Booster.

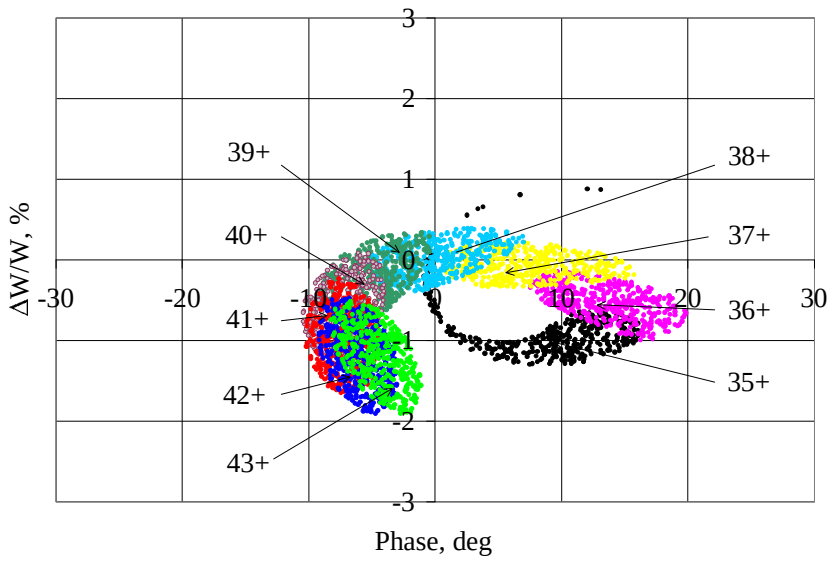

Figure 1: Longitudinal phase space plots of the accelerated multiple charge-state uranium beam exiting the booster.

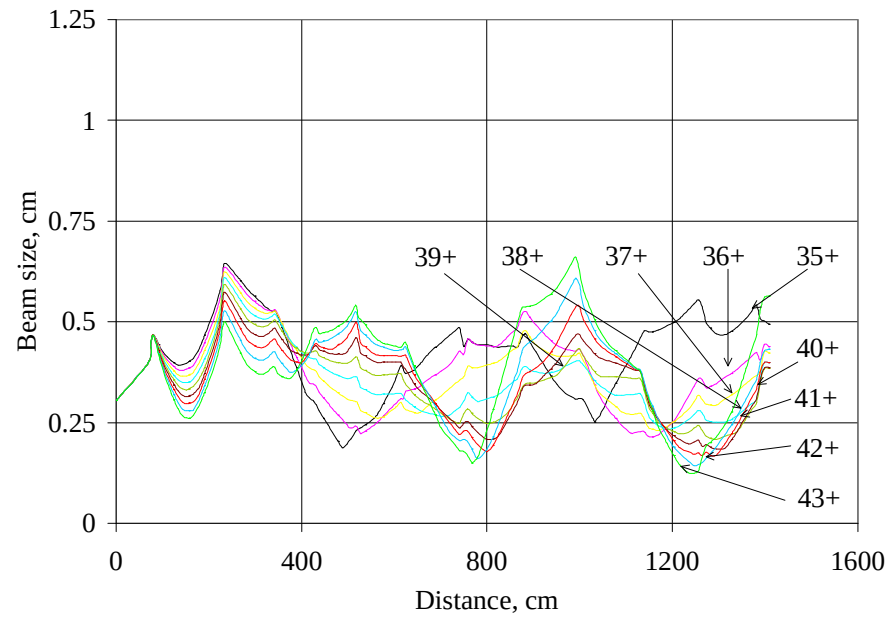

Figure 2: Transverse envelopes of the multiple charge-state uranium beam along the booster.

As is seen one can expect acceleration of most charge states produced after the stripping foil. The simulation of the transverse motion does not include the misalignment errors; and therefore the total effective emittance growth is negligible.

## 3 DESCRIPTION OF THE EXPERIMENT

The $^{238}U^{+26}$ beam from the ATLAS ECR-II ion source was accelerated to 286 MeV (~1.2 MeV/u) in the Injector Linac, and stripped in a 75 μg/cm$^2$ carbon foil 0.5 m before the 'Booster' linac as shown in Fig. 3. The beam energy was carefully measured by a resonant time-of-flight (TOF) system [3]. The ATLAS Booster was tuned using a $^{58}Ni^{+9}$ 'guide' beam from the ATLAS tandem injector whose velocity was matched to that of the stripped $^{238}U^{+38}$ and which has a similar charge-to-mass ratio. The synchronous phase for $^{238}U^{+38}$ was chosen to be –30°. Therefore the synchronous phase $\varphi_G$ required for the guide beam is given by [1]

$$\varphi_G = -\arccos\left[\frac{58 \cdot 38}{9 \cdot 238}\cos(-30°)\right] = -27° \ .$$

The synchronous phase in all 24 cavities of the booster is set by an auto-scan procedure using a silicon detector for beam energy measurements. Tuning of the focusing fields to get 100% transmission was accomplished with the guide beam prior to switching to the uranium mixed beam.

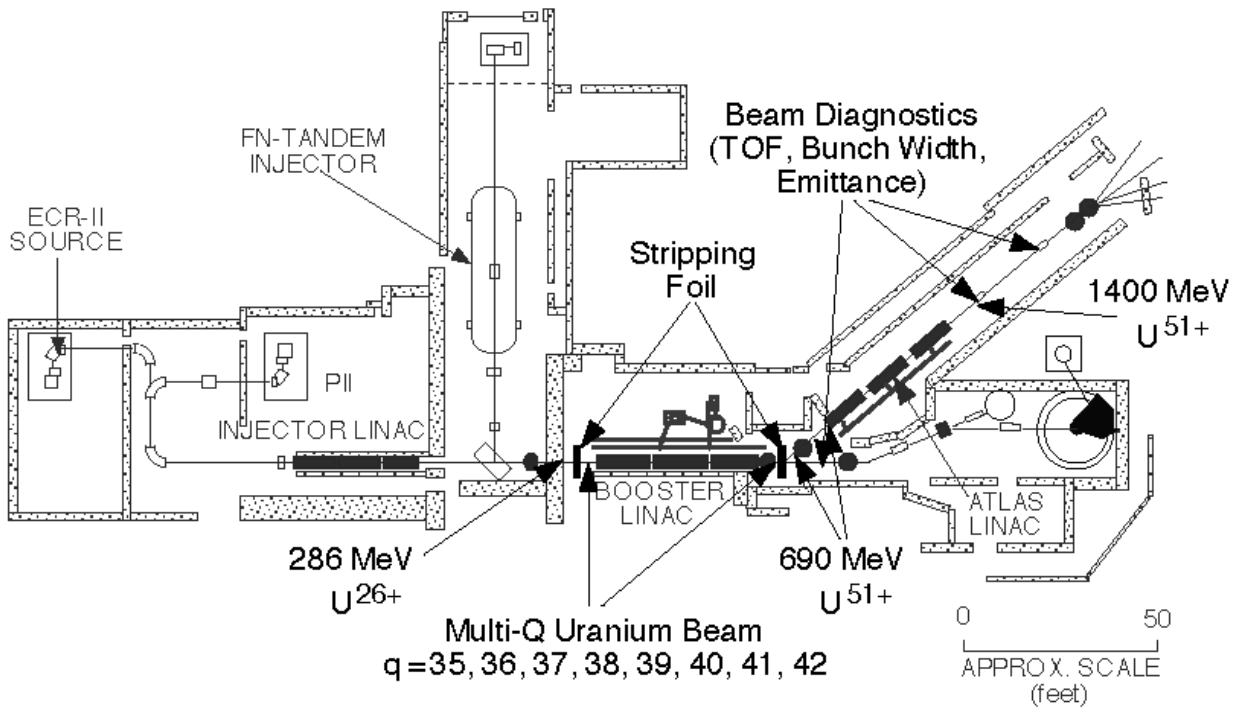

Figure 3: Layout of ATLAS linac.

After optimising the Booster linac and 40°-bend tune with the $^{58}Ni^{+9}$ guide beam, the stripped uranium beam was injected into the Booster. Magnet slits were used to cleanly select only the 38+ beam after the bending magnets and the uranium injection phase was matched to the guide beam's phase empirically based on maximum transmission through the system. Further tuning of the bunching system and last PII resonator made small adjustments to the uranium beam energy to better match the guide beam's velocity. After this tuning process, a 91% transmission of the multiple charge-state uranium beam was achieved. The transmission improved to 94% when a 10 mm aperture was inserted upstream of the stripping target. Fig. 4 compares the intensity distribution of the mixture of multiple charge-state uranium beams accelerated in the booster to the measured stripping distribution for the unaccelerated uranium. The difference in the distributions is caused mainly by poorer transmission of lower charge states through the booster. Also, some discrepancy is expected due to slightly different tuning of unaccelerated and accelerated beams and collimator slits in the 40° bend region.

The individual charge states then were analysed in the 40° bend region and sent to the ATLAS beam diagnostics area (see Fig. 3). The parameters of each selected charge state were carefully measured. Particularly the following beam parameters were measured:

- Transverse emittance (the value and ellipse orientation in phase space) by the help of quadrupole triplet gradient variation [4] and a wire scanner located 3.1 m apart.
- Average beam energy using the ATLAS TOF energy measurement system.
- Beam energy spread with the silicon detector measuring the bunch time width after a long drift space to the ATLAS diagnostics area.

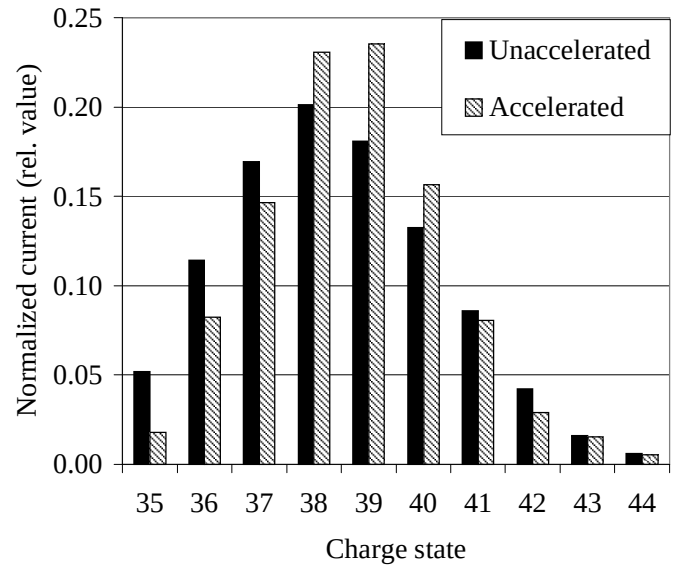

Figure 4: Comparison of intensity distributions for accelerated and unaccelerated multiple charge uranium beams.

Finally, the multi-charged uranium beam was stripped for the second time at the exit of the Booster and $^{238}U^{+51}$ was selected. The same beam parameters measurements were performed and the beam was further accelerated in the last section of ATLAS. As expected, the use of multi-charged uranium beam on the second stripper increased the intensity of double-stripped $^{238}U^{+51}$ beam. The double-stripped $^{238}U^{+51}$ was accelerated up to 1400 MeV and used for a scheduled experiment at ATLAS.

Basic results of these beam measurements are shown in Fig. 5-8. Figure 5 presents transverse beam profiles at the exit of the booster. The multiple charge-state uranium beam has a larger size compared to the guide beam. As was shown in ref. [1] misalignments of the focusing elements and effective emittance growth of a multi-charged beam compared to a single charge-state beam are the main source of the larger beam size. Such errors must be minimized in machines designed for the utilization of multi-charge beams. The results of individual transverse emittance measurements are presented in Table 1 and Figure 6. The horizontal emittance is less than the vertical due to the charge selection by the slits downstream of the bending magnet. Therefore only emittances in the vertical plane are shown. The double-stripped uranium beam $^{238}U^{+51}$ contains all information about the effective emittance of multi-charged beam output of the Booster. The effective emittance increases by a factor of 2 due to the misalignment errors of the solenoids in the Booster.

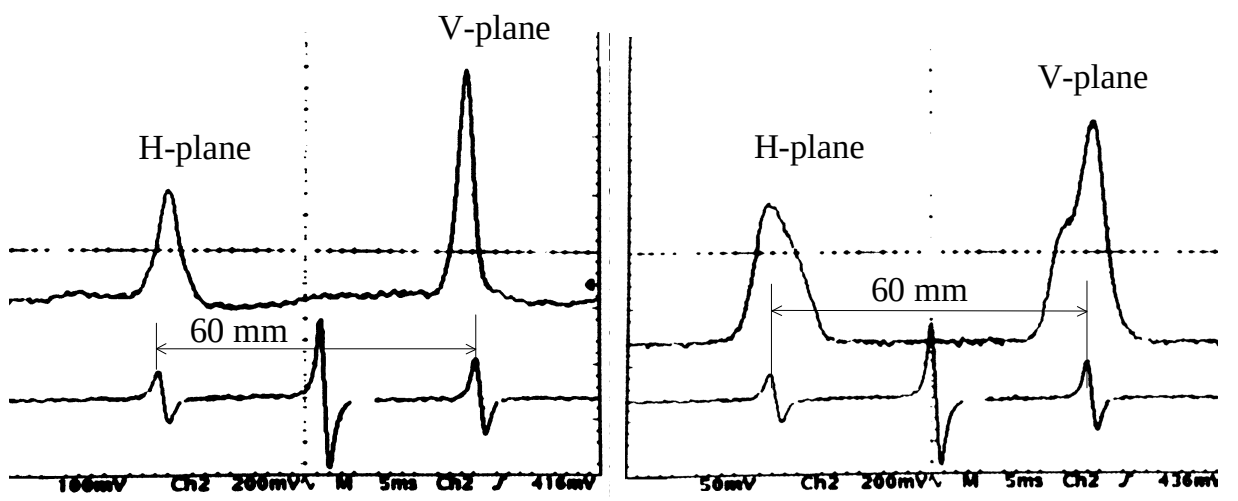

Figure 5: Transverse profiles at the exit of the booster for guide beam (left) and multi-Q uranium beam (right).

Average energy and the FWHM energy spread of individual charge states are shown in Fig. 7. For the simulation of energy spread, the input longitudinal emittance $\varepsilon_L = 2\pi\cdot$keV/u·nsec was assumed. The graphs show consistent behaviour of the energy spread as a function of charge state. There is some discrepancy with the average energy for the remote charge state which is, probably, caused by the longitudinal tuning errors of the SRF cavities. Certainly such tuning in high intensity machines should be done with high precision, but for routine operation of ATLAS it is not required. Even with these tuning errors, the average energy spread for three neighbouring charge states 37-39, similar to the beam energy distribution proposed for the RIA driver linac, is only 0.7%.

Table 1: Twiss parameters of single charge state beams at the exit of ATLAS for the vertical plane

| Uranium charge state | $\alpha_y$ | $\beta_y$, mm/mrad | $\varepsilon_{y,\,normalized}$, $\pi$·mm·mrad |
|---|---|---|---|
| 36+ | 0.72 | 12.66 | 0.94 |
| 37+ | 0.48 | 8.08 | 1.24 |
| 38+ | 0.06 | 10.17 | 1.11 |
| 39+ | 0.45 | 7.60 | 1.34 |
| 40+ | 0.54 | 9.22 | 1.03 |
| 41+ | -0.18 | 9.20 | 0.89 |
| 51+ | 0.60 | 9.00 | 2.69 |

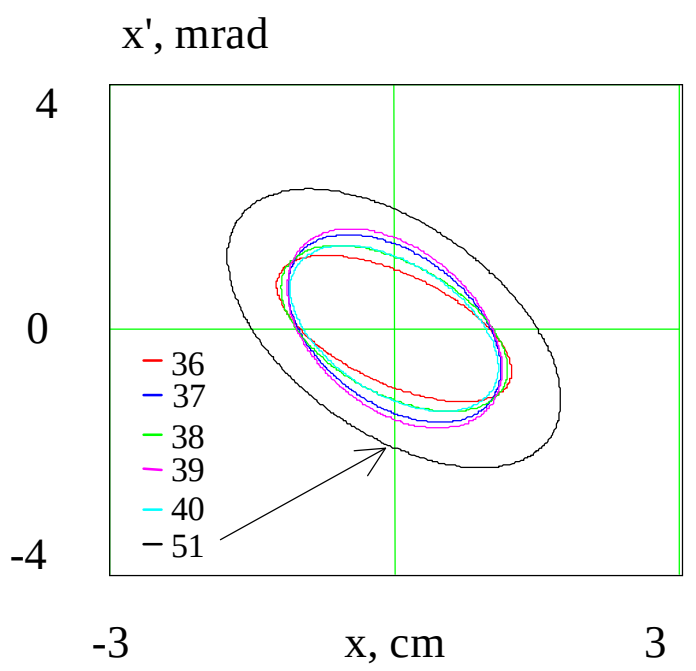

Figure 6: Vertical phase space ellipses of single charge-state beams. The black ellipse corresponds to double-stripped $U^{51+}$.

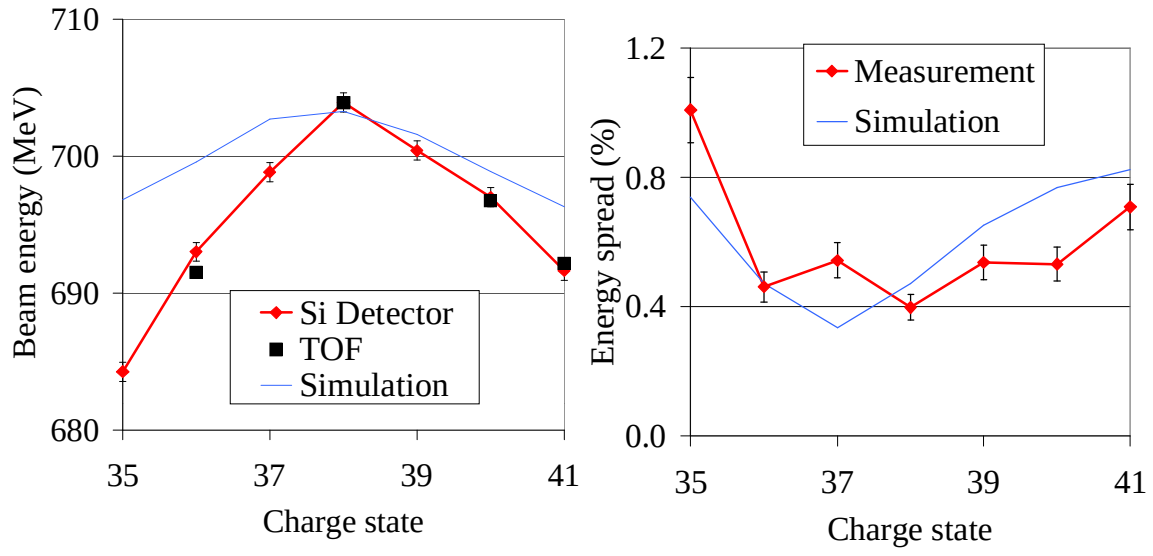

Figure 7: Beam energy (left) and the FWHM energy spread (right) of individual charge states.

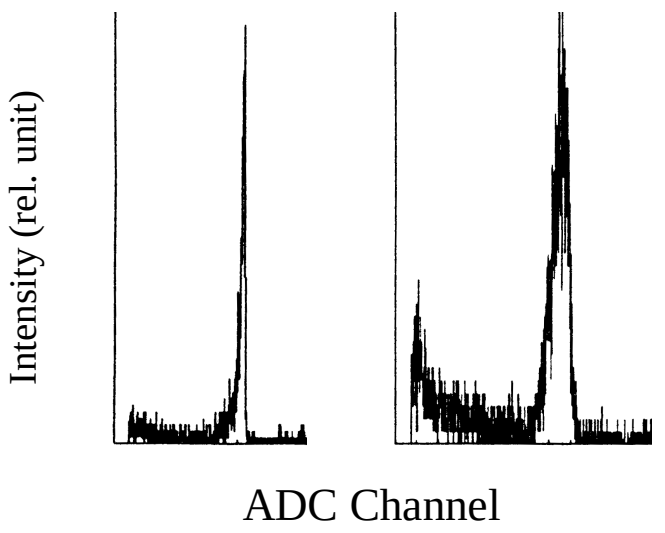

Figure 8: Bunch time width for a single charge state $^{238}U^{+38}$ beam (left) and a double-stripped $^{238}U^{+51}$ beam (right).

Figure 8 presents bunch time width measurements of beams transported to the ATLAS diagnostic area. The left spectrum corresponds to charge state 38+ (before the second stripping) and the right spectrum belongs to double-stripped uranium beam $^{238}U^{+51}$. The low intensity background events are mostly due to detector system background. The energy spectra at FWHM obtained from these measurements are 0.4% for charge state 38+ and 1.3% for charge state 51+. So, the second stripping of the multi-charged beam produced a 3 times larger energy spread.

# CONCLUSION

The results of this test are consistent with the simulation and show that multi-charged beam acceleration can substantially increase the intensity of heavy-ion beams. A medium-energy high-power machine, such as the RIA driver linac, can be designed to utilize multi-Q beams if unwanted charge states after each stripping target are cleaned by a corresponding magnetic system. For low-intensity and low-energy linacs, such as ATLAS, the double-stripped heavy-ion beam can be used to obtain higher beam energy while providing more intensity than with single charge-state acceleration.

Work supported by the U. S. Department of Energy under contract W-31-109-ENG-38